# ModelChain: Decentralized Privacy-Preserving Healthcare Predictive Modeling Framework on Private Blockchain Networks

Tsung-Ting Kuo, PhD[1] and Lucila Ohno-Machado, MD, PhD[1,2]
[1]UCSD Health System Department of Biomedical Informatics, University of California San Diego, La Jolla, CA
[2]Division of Health Services Research & Development, VA San Diego Healthcare System

**Abstract**
Cross-institutional healthcare predictive modeling can accelerate research and facilitate quality improvement initiatives, and thus is important for national healthcare delivery priorities. For example, a model that predicts risk of re-admission for a particular set of patients will be more generalizable if developed with data from multiple institutions. While privacy-protecting methods to build predictive models exist, most are based on a centralized architecture, which presents security and robustness vulnerabilities such as single-point-of-failure (and single-point-of-breach) and accidental or malicious modification of records. In this article, we describe a new framework, ModelChain, to adapt Blockchain technology for privacy-preserving machine learning. Each participating site contributes to model parameter estimation without revealing any patient health information (i.e., only model data, no observation-level data, are exchanged across institutions). We integrate privacy-preserving online machine learning with a private Blockchain network, apply transaction metadata to disseminate partial models, and design a new proof-of-information algorithm to determine the order of the online learning process. We also discuss the benefits and potential issues of applying Blockchain technology to solve the privacy-preserving healthcare predictive modeling task and to increase interoperability between institutions, to support the Nationwide Interoperability Roadmap and national healthcare delivery priorities such as Patient-Centered Outcomes Research (PCOR).

**Introduction**
Cross-institution interoperable healthcare predictive modeling can advance research and facilitate quality improvement initiatives, for example, by generating scientific evidence for comparative effectiveness research,[1] accelerating biomedical discoveries,[2] and improving patient-care.[3] For example, a healthcare provider may be able to predict certain outcome even if her institution has few or none related patient records. A predictive model can be "learned" (i.e., its parameters can be estimated) from data originating from the other institutions. However, improper data disclosure could place sensitive personal health information at risk. To protect the privacy of individuals, several algorithms (such as GLORE,[4] EXPLORER,[5] and VERTIGO[6]) have been proposed to conduct predictive modeling by transfer of partially-trained machine learning models instead of disseminating individual patient-level data. However, these state-of-the-art distributed privacy-preserving predictive modeling frameworks are centralized (i.e., require a central server to intermediate the modeling process and aggregate the global model),[4–6] as shown in Figure 1(a). Such a client-server architecture carries the following risks:

- *Institutional policies*. For example, a site may not want to cede control to a single central server.[7]

- *Single-point-of-failure*.[8,9] For example, if the central server is shut down for maintenance, the whole network stops working. Furthermore, if the admin user account of the central server gets compromised, the entire network is also under the risk of being compromised.[7]

- *Participating sites cannot join/leave the network at any time*.[10] If any site joins or leaves the network for a short period of time, the analysis process is disrupted and the server needs to deal with the recovering issue. A new site may not participate in the network without the authentication and reconfiguration on the central server.[8,9]

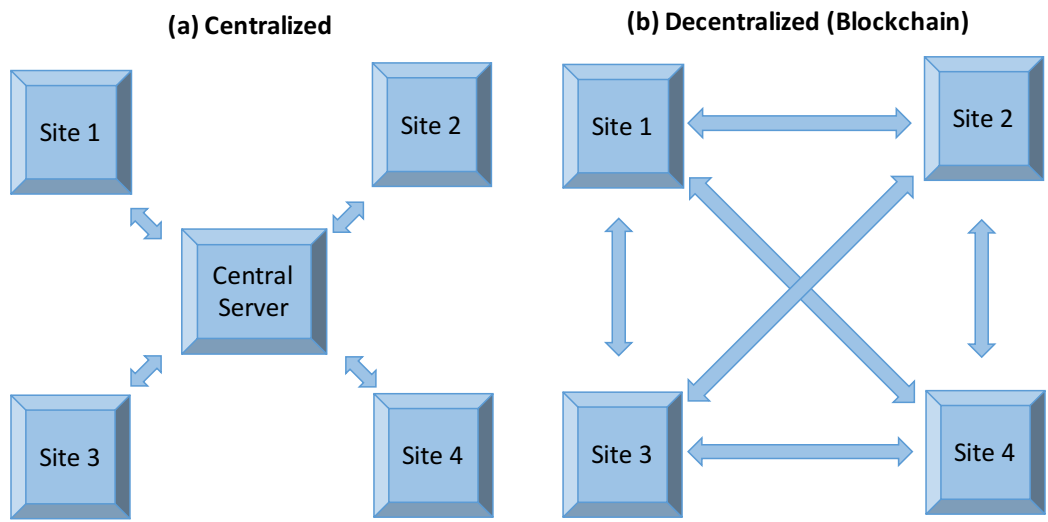

**Figure 1.** (a): Centralized topology. (b): Decentralized topology (Blockchain).

- *The data being disseminated and the transfer records are mutable*. An attacker could change the partial models without being noticed.[7] The transfer records may also be modified so that no audit trail is available to identify such malicious change of data.[11,12]

- *The client-server architecture may present consensus/synchronization issues on distributed networks*. Specifically, the issue is the combination of two problems: the Byzantine Generals Problem,[13] in which the participating sites need to agree upon the aggregated model under the constraint that each site may fail due to accidental or even malicious ways,[7] and the Sybil Attack Problem,[14] of which the attacker comprises a large fraction of the seemingly independent participants and exerts unfairly disproportionate influence during the process of predictive modeling.[7,15]

To address the abovementioned risks, one plausible solution is to adapt the *Blockchain* technology (in this article, we use "Blockchain" to denote the technology, and "blockchain" to indicate the actual chain of blocks).[7,9–12,15–20] A Blockchain-based distributed network has the following desirable features that make it suitable to mitigate the risks of centralized privacy-preserving healthcare predictive modeling networks. First, Blockchain is by design a decentralized (i.e., a peer-to-peer, non-intermediated) architecture (Figure 1(b)); the verification of transactions is achieved by majority *proof-of-work* voting.[17] Each institution can keep full control of their own computational resources. Also, there is no risk of single-point-of-failure.[8,9] Second, each site (including new sites) can join/leave the network freely without imposing overhead on a central server or disrupting the machine learning process.[8–10] Finally, the proof-of-work blockchain provides an immutable audit trail.[7,11,12] That is, changing the data or records is very difficult; the attacker needs to redo proof-of-work of the target block and all blocks after it, and then surpass all honest sites. As shown by Satoshi Nakamoto,[17] the inventor of Blockchain and Bitcoin, given that the probability that an honest node finds the next block is *larger* than the probability that an attacker finds the next block, the probability the attacker will ever catch up drops exponentially as the number of the blocks by which the attacker lags behind increases. This is also the reason why the Blockchain mechanism also solves the relaxed version of Byzantine Generals Problem and the Sybil Attack Problem,[9,15,18,20] as formally proved by Miller et al.[18]

Although Blockchain provides the abovementioned security and robustness benefits, a reasonable approach to integrate Blockchain with the privacy-preserving healthcare predictive modeling algorithms is yet to be devised. In this article, we propose *ModelChain*, a private-Blockchain-based privacy-preserving healthcare predictive modeling framework,to combine these two important technologies.

First, we apply privacy-preserving *online* machine learning algorithms on blockchains. Intuitively, the incremental characteristic of online machine learning makes it feasible for peer-to-peer networks like Blockchain. Then, we utilize *metadata* in the transactions to disseminate the partial models and other meta information (i.e., flag (which indicates the type of action) of the model, hash of the model, and error of the model), and thus integrate private blockchains (i.e., the network is available only for participating institutions) with privacy-preserving online machine learning. Finally, we design a new *proof-of-information* algorithm on top of the original proof-of-work consensus protocol, to determine the order of the online machine learning on blockchains, aiming at increasing efficiency and accuracy. The basic idea of proof-of-information is similar to the concept of Boosting:[21–25] the site that contains data that *cannot* be predicted accurately using a current partial model contains *more information* to improve the model, and thus that site should be assigned a higher priority to be chosen as the next model-updating site. We start with the best model to prevent error propagation, choose the site with highest error for current model to update the model, and repeat the process to update the model until a site cannot find any other site with higher error to update the model. In this case, we consider the model as the *consensus model*.

ModelChain can advance the following interoperability needs stated in the Nationwide Interoperability Roadmap[26] of the Office of the National Coordinator for Health Information Technology (ONC):

- *"Build upon the existing health IT infrastructure."* ModelChain exploits the existing healthcare data in Clinical Data Research Networks (CDRNs) such as the Patient-centered SCAlable National Network for Effectiveness Research (pSCANNER)[27], which is one of the Clinical Data Research Networks (CDRNs) in the PCORI-launched PCORnet[89,90] and includes three networks: VA Informatics and Computing Infrastructure (VINCI),[28,29] University of California Research eXchange (UCReX),[30] and SCANNER.[31,32] With the support of the Blockchain backbone, ModelChain can leverage all existing patient data storage infrastructures, while improving the healthcare prediction power for every site.

- *"Maintain modularity."* Comparing to traditional client-server architecture, ModelChain inherits the peer-to-peer architecture of Blockchain, allowing each site to remain modular while interoperating with other sites. Also, each site has control about how its data are accessed (instead of ceding control to the central server), thus can keep up with institutional policies. Moreover, Blockchain provides the native ability to automatically coordinate the joining or leaving of each site, further improving the independence and modularity for the participating institutions.

- *"Protect privacy and security in all aspects of interoperability."* ModelChain is designed to provide a secure, robust and privacy-preserving interoperability platform. Specifically, Blockchain increases the security by avoiding single-point-of-failure, proving immutable audit trails, and mitigating the Byzantine Generals and the Sybil Attack problems, while preserving the privacy by exchanging zero patient data during the predictive modeling process.

The expected benefits of ModelChain can also be linked to the stated objectives of Patient-Centered Outcomes Research (PCOR)[33–35] defined by the Patient-Centered Outcomes Research Institute (PCORI).[36–38]

## Related Work

### Privacy-preserving predictive modeling

Cross-institutional healthcare predictive modeling and machine learning can accelerate research and facilitate quality improvement initiatives. However, improper information exchange of biomedical data can put sensitive personal health information at risk. To protect the privacy of individuals, many algorithms[4–6,39–46] have been proposed to conduct predictive modeling by transfer of partially-trained machine learning models, instead of disseminating individual patient data. For example, GLORE[4] built logistic regression models with horizontally partitioned data, VERTIGO[6] dealt with vertically partitioned data, and WebDISCO[47] constructed Cox proportional hazards model on horizontally partitioned data.

Among these distributed privacy-preserving machine learning algorithms, EXPLORER[5] and the Distributed Autonomous Online Learning[45] are "online" machine learning algorithms of which models can be updated in a sequential order (as opposed to the other "batch" algorithms). Such an online machine learning algorithms are similar to our proposed ModelChain that updates models on Blockchain sequentially.

However, all these machine learning algorithms, which either update the models in a batch or online fashion, relied on a centralized network architecture that may suffer from security risks such as a single-point-of-failure. In contrast, ModelChain is built on top of Blockchain, which is a decentralized architecture and can provide further security/robustness improvement (e.g., immutable audit trails).

Another related area covers distributed data-parallelism machine learning algorithms,[48] such as Parameter Server[49–52] or compute models using the MapReduce[53–56] technology. Nevertheless, they mainly focus on the parallelization algorithms to speed-up the computation process, instead of aiming at privacy-preserving data analysis,[4] and thus are different from our method.

### Blockchain technology for crypto-currency applications

Blockchain was first proposed as a *proof-of-work* consensus protocol implementation of peer-to-peer timestamp server on a decentralized basis in the famous *Bitcoin* crypto-currency.[17] Specifically, an electronic coin (e.g., Bitcoin) is defined as a chain of transactions. A block contains multiple transactions to be verified, and the blocks are chained (i.e., "blockchain") using hash functions to achieve the timestamp feature.

Then, each site "mines" blocks (to confirm the transactions) by solving a difficult hashing problem (i.e., "proof-of-work"). That is, each block contains an additional counter (i.e., "nonce") as one of the inputs of the hash function, and the nonce is incremented until the hashed value contains specified leading zero bits (then the work is "proofed").[17] The first site that successfully satisfies the proof-of-work (and thus has the "decision power"[57]) verifies the transactions and adds the confirmed block at the end of the blockchain, and the block is confirmed and is considered "immutable";[17] if any attacker wants to change a block, all the blocks after it would also require to be recomputed (because each block is computed using the hash of the previous block in the chain). Given the assumption that honest computational sites (i.e., computational power) are larger than malicious sites, the probability that the attacker can recompute and modify a block is extremely small (especially when the attacker has already lagged behind for many blocks).[17]

Such a proof-of-work design can also be regarded as majority voting (i.e., one-CPU-one-vote); the longest chain (invested with the heaviest proof-of-work effort) represents the majority decision, and thus no trusted central authority (i.e., "mint") is required to prevent the double-spending problem (i.e., the transactions are validated by the longest chain - the majority of the sites). Several recent researches provide detailed analyses of the Blockchain consensus protocol in terms of its ability to resist attacks.[17,20,58–61]

After Bitcoin, several alternatives have also been proposed (alternative blockchains, or "altchains"), such as Colored coins[63] (a protocol to support Bitcoin in different "colors" as different crypto-currencies) and Side-chains[64,65] (a protocol to allow Bitcoin to be transferred between multiple blockchain networks).

Also, several protocol have been proposed on top of Bitcoin's proof-of-work to increase the difficulty of developing a "Bitcoin monopoly", such as proof-of-stake[57,66,67] (in which the "decision power" is based on the ages of the owned bitcoins; the site with the largest "stake" can confirm and add the new block to the blockchain) and proof-of-burn[65,68] (in which the "decision power" is based on the destroying of the owned bitcoins; the site that is willing to destroy the largest number of its bitcoins can confirm and add the new block to the blockchain). In this article, we propose a proof-of-information algorithm on top of the proof-of-work, to provide "decision power" (i.e., privilege to update the online machine learning model) to the site with the highest expected amount of information.

Blockchain technology for non-financial and healthcare applications
Blockchain was created for financial transactions, but it is also a new form of a distributed database, because it can store arbitrary data in the transaction metadata (the metadata has been an official Bitcoin entity since 2014).[7,10,69,70] The original Bitcoin only supports 80 bytes of metadata (via OP_RETURN), but several implementations of Blockchain support a larger metadata size. For example, MultiChain[10] supports adjustable maximum metadata size per transaction. Another example is BigchainDB,[7] which is built on top of a big data database RethinkDB[71] and thus has no hard limit on the transaction size. Here, we utilize the transaction metadata to disseminate the partially trained online machine learning model (and the meta information of the model) among participating sites. Such Blockchain-based distributed database is also known as Blockchain 2.0, including technologies such as *smart properties* (the properties with blockchain-controlled ownership) and *smart contracts* (computer programs that manage smart properties).[63,64,72–80] One of the most famous Blockchain 2.0 system is Ethereum,[73,78] a decentralized platform that runs smart contracts. Ethereum has a built-in Turing-complete programming language that supports loop computation, which is not provided by the Bitcoin scripting language.[73,78] In the context of a distributed database, smart properties are data entries, and smart contract are stored procedures. Our proof-of-information algorithm may be implemented using Blockchain 2.0 technologies as well, with smart properties being partial models, and smart contracts being the algorithms to update and transfer the partial models.

Recently, the concept of Blockchain 3.0 has been proposed to indicate applications beyond currency, economy, and markets.[75] One of the most important application is the adaption of Blockchain technology to the healthcare system. For example, Irving et al. evaluated the idea of using the blockchain as a distributed tamper-proof public ledger, to provide proof of pre-specified endpoints in clinical trial;[81] McKernan proposed to apply decentralized blockchain to store genomic data;[82] and Jenkins et al. discussed a bio-mining framework for biomarkers with a multi-resolution blockchain to perform multi-factor authentication and thus increase data security.[83] There are also studies that propose to use Blockchain to store electronic health records,[84,85] or to record health transactions.[86] However, to the best of our knowledge, we are the first to propose the adoption of Blockchain to improve the security and robustness of privacy-preserving healthcare predictive modeling.

**The ModelChain Framework**
In ModelChain, we apply privacy-preserving *online* machine learning algorithms on blockchains. Intuitively, the incremental characteristics of online machine learning is feasible for peer-to-peer networks like Blockchain. It should be noted that any online learning algorithm, such as EXPLORER[5] or Distributed Autonomous Online Learning,[45] can be adapted in our framework.

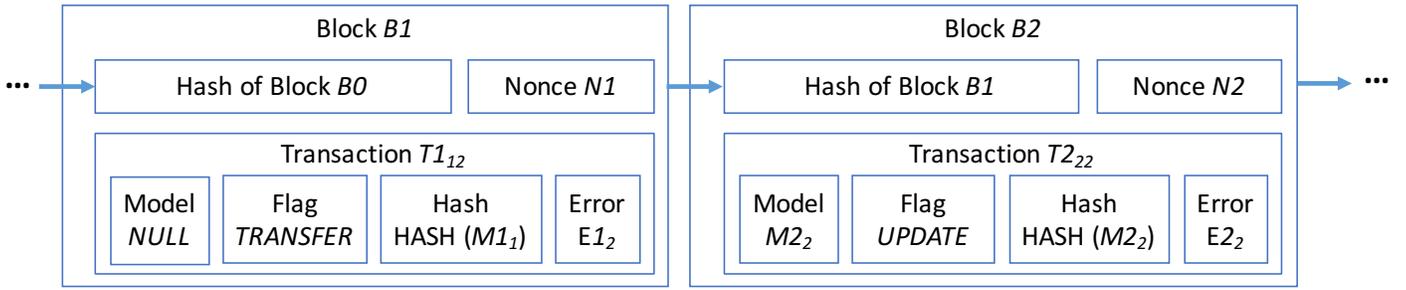

**Figure 2.** An example of ModelChain. Each block represents a timestamp, containing only one transaction. Each transaction contains a model, flag (action type) of the model, hash of the model, and error of the model.

Next, we utilize the metadata in the transactions to disseminate the partial models and the meta information (i.e., flag of the model, hash of the model, and error of the model) to integrate privacy-preserving online machine learning with a private Blockchain network (Figure 2). There are four types of flag in ModelChain: INITIALIZE, UPDATE, EVALUATE, and TRANSFER, which indicates the action a site has taken to a model (e.g., INITIALIZE = the site initialized the model). We include the hash of the model to save storage spaces (i.e., only UPDATE transactions include both model and hash of model; all other type of transactions only include hash of the model (and model = NULL) to reduce the size of blockchain). In a transaction, both the amount of the transactions and the transaction fees are set to be zero. Also, in this private Blockchain network, no block mining reward is provided. The incentive for each site to mine blocks and verify transactions is the improved accuracy of the predictive model using cross-institution data in a privacy-preserving manner. Besides, a block can only contain one transaction (so each transaction has a unique timestamp). The private blockchain containing all blocks of transactions can be regarded as a distributed database (or data ledger) that every site can read and write to. We then use this Blockchain-based private distributed database as a basis of the proof-of-information algorithm. Finally, we designed a new proof-of-information algorithm on top of the original proof-of-work consensus protocol, to determine the order of the online machine learning on blockchains, aiming at increasing efficiency and accuracy. The basic idea is similar to the concept of Boosting:[21–25] the site which contains data that *cannot* be predicted accurately using current partial model probably contains more information to improve the model than other sites, and thus that site should be assigned a higher priority to be chosen as the next model-updating site.

A running example of the proof-of-information algorithm is shown in Figure 3. Suppose there are four participating sites that would like to train a privacy-preserving online machine model on the private Blockchain network. Assume $Mt_s$ = model at time $t$ on site $s$, $Et_s$ = error at time $t$ on site $s$. In the initialization stage ($t$ = 0), each site trains their own model using their local patient data, and the model with lowest error (Site 1 with $E0_1$ = 0.2 in our example) is selected as the initial model. The reason to choose the *best* model is to prevent the propagation of error. Conceptually, we regard $M0_1$ is "transferred" from Site 1 to Site 1 itself. Then, the selected model ($M0_1$) is submitted to Site 2, 3 and 4.

Next ($t$ = 1), each site evaluates the model $M1_1$ (which is the same as $M0_1$) using their local data. Suppose Site 2 has the highest error ($E1_2$ = 0.7). Given that the data in Site 2 is the most unpredictable for model $M1_1$, we assume that Site 2 contains the richest information to improve $M1_1$. Therefore, Site 2 wins the "information bid", and the model $M1_1$ is now transferred to Site 2 within the block $B1$ (with amount = 0 and transaction fee = 0) shown in Figure 2. It should be noted that the Blockchain protocol requires every site to submit every transaction to each other for verification. Therefore, $M1_1$ is actually submitted from Site 1 to every site. However, since Site 2 wins the "information bid", we conceptually regard that $M1_1$ is "transferred" from Site 1 to Site 2, in the sense that only Site 2 can update $M1_1$ using the local patient data in Site 2.

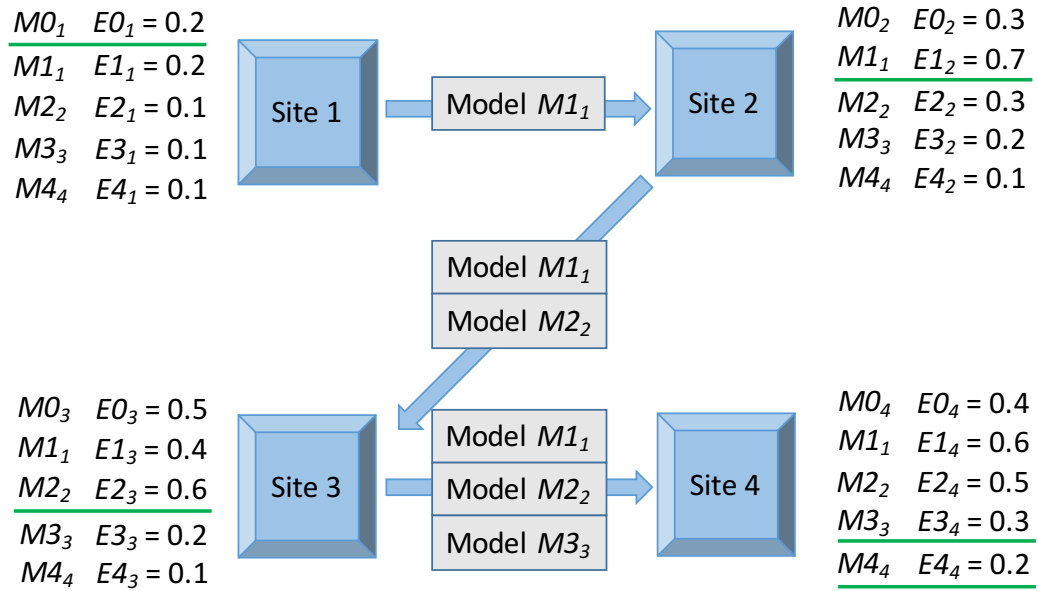

**Figure 3.** An example of the proof-of-information algorithm. $Mt_s$ = model at time $t$ on site $s$, $Et_s$ = error at time $t$ on site $s$. The model/error with green underline is the selected one at that timestamp (at each $t$, only one model/error is selected).

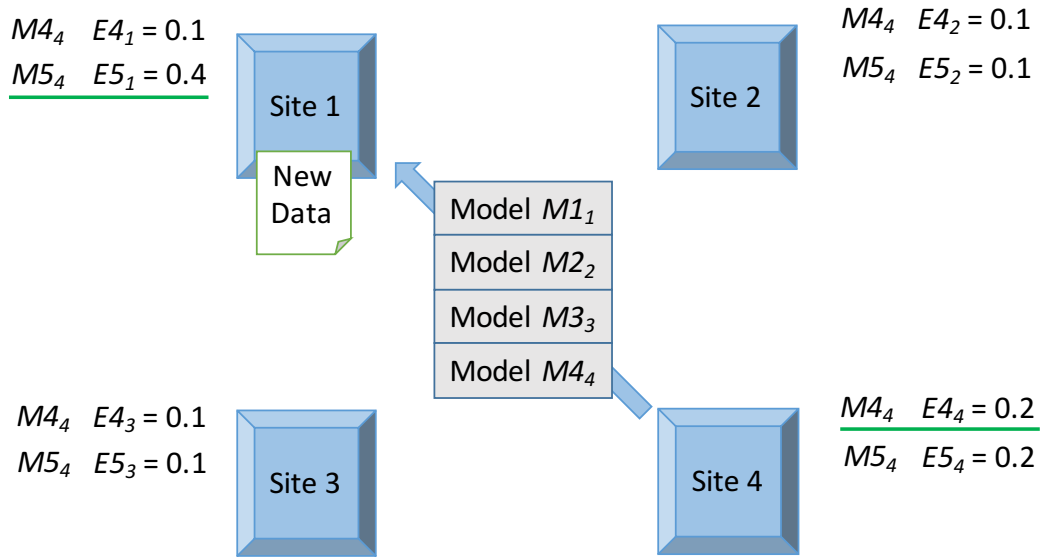

**Figure 4.** An example of the proof-of-information algorithm for new data. Suppose the current ($t = 4$) consensus model is $M4_4$, and the new data is in Site 1. $Mt_s$ = model at time $t$ on site $s$, $Et_s$ = error at time $t$ on site $s$. The model/error with green underline is the selected one at that timestamp.

Then ($t = 2$), Site 2 updates the online machine learning model as $M2_2$ (within the block $B2$ shown in Figure 2). Again, Site 2 send $M2_2$ to all other sites, and the site with highest error (or richest information) wins the "information bid" to update the model locally (Site 3 in our example). This process repeats until a site updates the model and finds that itself has the highest error than all other sites. For example, when $t = 3$, Site 4 has the highest error (0.3) and thus wins the bid to update the model; but when $t = 4$, Site 4 still has the highest error (0.2) using the updated model. Therefore, the process does not need to continue; we regard the model as consensus and the online machine learning process stops, with $M4_4$ as the consensus model.

**Algorithm 1.** Proof-of-Information-Iteration. This is the core algorithm to determine the order of decentralized privacy-preserving online machine learning.

**Input**: this site $S$, polling time period $\Delta$, waiting time period $\Theta$
**Output**: the latest online machine learning model $M$

1    For every time period $\Delta$ check the block chain
2      If (there are new models (flag = UPDATE) in the block chain)
3        Retrieve the latest model $M_C$ (generated by site $C$) and current largest error $E_C$ from the block chain
4        Set $M = M_C$
5        Evaluate $M_C$ on the data in $S$ and compute the error $E$
6        Create a transaction from $S$ to $S$ itself with flag = EVALUATE, model = NULL, hash = HASH ($M_C$), and error = $E$
7      If (the model $M_C$ (flag = TRANSFER) is transferred from $C$ to $S$)
8        Update $M_C$ using the data in $S$ to generate the new model $M_S$ and new error $E_S$
9        Set $M = M_S$
10        Create a transaction from $S$ to $S$ itself with flag = UPDATE, model = $M_S$, hash = HASH ($M_S$), and error = $E_S$
11        Wait for specific time period $\Theta$ and collect all errors (with flag = EVALUATE) from other sites
12        If ($E_S$ is not larger than all errors)
13          Identify the site $L$ with the largest error $E_L$
14          Create a transaction from $S$ to $L$ with flag = TRANSFER, model = NULL, hash = HASH ($M_S$), and error = $E_L$

In the case any site adds new data, we do not need to re-train the whole model. Instead, we use the proof-of-information again to determine whether we should update the model using the new data. As illustrated in Figure 4, suppose the current ($t = 4$) consensus model is $M4_4$, and the new data is in Site 1. In time $t = 5$, Site 1 use the updated data (including both old and new data) to evaluate model $M5_4$ (which is the same as $M4_4$) and realized that the error $E5_1 = 0.4$ is larger than current updating site (i.e., Site 4 with $E5_4 = 0.2$). Therefore, Site 1 wins the "information bid" again, and the model $M5_4$ is now transferred to Site 1 to be updated. Then, the same process shown in Figure 3 repeated until identifying the consensus model. Note that if the error $E5_1$ was lower than $E5_4$, we consider that the new data does not bring enough information to the model $M5_4$, thus no transfer and update are required. The similar mechanism can be used for a new site (that is, new site can be treated as a site where all data are new).

Another situation to be considered is the case in which a site leaves the private Blockchain network. Based on the Blockchain mechanism, we do not need to deal with the site-leaving situation. If the site leaves while not updating the model (e.g., Site 2 at time $t = 5$ in Figure 4), then we can simply ignore the departure; the site can join at any time under the Blockchain mechanism. On the other hand, if the site leaves while updating the model (e.g., Site 1 at time $t = 5$ in Figure 4), we can still ignore it. This is because such a model transferring is only conceptual; the model in the blockchain is not updated (i.e., the latest model $M5_4$ is at the end of the blockchain), until Site 1 completes the model update. Therefore, each site in the network can still use the latest model locally, and once Site 1 comes back to the network, it is treated as a new site so it can continue the model update process.

The detailed proof-of-information algorithms are shown in Algorithm 1, 2 and 3. Algorithm 1 is the core of the proof-of-information algorithm, which determines the order of learning and repeats the learning process until the consensus model is found. For a new Blockchain network, each site executes Algorithm 2, which in turn executes Algorithm 1. For a new participant to an existing network, or an existing participant with new data, the site executes Algorithm 3, which also leverages Algorithm 1 to learn the consensus model.

**Algorithm 2.** Proof-of-Information-Initialization. This is the main algorithm for a new network (i.e., all participating sites are new).

**Input**: this site $S$, polling time period $\Delta$, waiting time period $\Theta$, total number of participating sites $N$
**Output**: the latest online machine learning model $M$

1     Learn model $M_S$ on the data in $S$ and compute the error $E$
2     Set $M = M_S$
3     Create a transaction from $S$ to $S$ itself with flag = *INITIALIZE*, model = *NULL*, hash = HASH ($M_S$), and error = $E_S$
4     Wait until errors (flag = *INITIALIZE*) from all $N$ sites on the network are received
5     if ($E$ is the smallest error among all errors)
6        Create a transaction from $S$ to $S$ with flag = *TRANSFER*, model = *NULL*, hash = HASH ($M_S$), and error = $E_S$
7     Set $M$ = Proof-of-Information-Iteration ($\Delta$, $\Theta$)

**Algorithm 3.** Proof-of-Information-New. This is the main algorithm for new participating site, or the existing site with newly available data.

**Input**: this site $S$, polling time period $\Delta$, waiting time period $\Theta$
**Output**: the latest online machine learning model $M$

1     Retrieve the latest model $M_C$ (generated by site $C$) and current largest error $E_C$ from the block chain
2     Set $M = M_C$
3     Evaluate $M_C$ on the data in $S$ and compute the error $E$
4     if ($E > E_C$)
5        Create a transaction from $C$ to $S$ with flag = *TRANSFER*, model = *NULL*, hash = HASH ($M_C$), and error = $E$
6     Set $M$ = Proof-of-Information-Iteration ($\Delta$, $\Theta$)

It should be noted that Algorithm 1 is actually a "daemon" service that is always watching the blockchain to check if any newly updated model is available. Therefore, although at times the consensus learning process in Algorithm 1 may pause due to the confirmation of the consensus model, Algorithm 1 keeps running and never stops (unless the site running it leaves the network, or the site has new data and would like to stop it to run Algorithm 3 instead). This mechanism of watching and responding events also suggests that our proposed proof-of-information algorithm may be implemented using Blockchain 2.0 technologies such as *smart properties* and *smart contracts*,[63,64,72–80] and be automatically executed at every site in the private network. That is, we can regard the partial models as smart properties, and realize the proof-of-information algorithm using smart contracts on each site to turn them to autonomous machines.

**Discussion**
Under the context of distributed privacy-preserving healthcare predictive modeling, Blockchain technology enables the following benefits: decentralization, freely joining/leaving, immutable records, and security improvements to deal with the Byzantine Generals and Sybil Attack Problems. However, there are also intrinsic limitations to Blockchain. First, confidentiality is not fully preserved: any site can trace all of the transactions and hence the error at each site (although the transactions are anonymous). Second, the transaction time can be long because of the proof-of-work computation (e.g., the average transaction time for Bitcoin is 10 minutes). Finally, it is vulnerable to the "51% attack",[10,19,73] which happens when there are more malicious sites than honest sites in the network.

Nevertheless, these limitations are less important for privacy-preserving healthcare predictive modeling. First, the main goal is to learn a better model using all patients' data without transferring personal protected health information. Second, the machine learning process itself may take a long time, especially for participating institutions with large patient data. In comparison, the transaction time is relatively small and is not a serious issue. Finally, the participating sites are healthcare institutions in a private Blockchain network, so the risk of the "51% attack" is minimal. One potential issue for the proof-of-information algorithm is that it might run too many iterations (i.e., model transferring transactions) without finding the "best" consensus predictive model. To resolve this issue, we can stop the algorithm if the error reaches a certain predefined threshold (the model is good enough), add a time-to-leave counter to limit the maximum number of iterations (the model is old enough), or apply both criteria (the model is either good enough or old enough). This way we can prevent ModelChain from running forever and consuming unnecessary computational power.

We also address implementation aspects of ModelChain as follows. First, we might consider setting the polling time period $\Delta$ and the waiting time period $\Theta$ in Algorithm 1, 2 and 3 based on the timeliness of model updating and the accuracy requirement of the models, the computational capability of the sites, and the underlying network environment. For example, if the data in each site update quickly, we might want to reduce the polling time period; if we require models with higher predictive power, then the waiting time period should be longer to find the best model. However, if the computational power or the network speed between sites are limited, it might not be feasible with low polling periods. Similarly, if we need more timely-updated models, shorter waiting time period would be preferable. Therefore, all the above factors should be considered to determine the best time period parameters. Second, the size of metadata should be considered, because every site stores a copy of the whole blockchain. Take EXPLORER[5] for example, the total model size is ($m * (m + 1)$), where $m$ is the size of the features. Suppose the model we want to construct has 1,000 features, the model size would be $(1{,}000 * (1{,}000 + 1)) * 64$ bits $\sim= 8$MB, which is exactly the default maximum metadata size of the MultiChain implementation.[10] Therefore, we consider the ModelChain framework reasonable in terms of metadata size. Finally, to further improve security, we can encrypt the transaction metadata that contains the model information, transmit the data via a virtual private network (VPN), and deploy ModelChain on private Health Insurance Portability and Accountability Act (HIPAA)-certified cloud computing environment such as integrating Data for Analysis, Anonymization, and Sharing (iDASH).[87,88]

**Conclusion**
The capability to securely and robustly construct privacy-preserving predictive model on healthcare data is essential to achieve the stated objectives in support of the Nationwide Interoperability Roadmap and national healthcare delivery priorities such as Patient-Centered Outcomes Research (PCOR). In this article, we proposed to improve the security and robustness of distributed privacy-preserving healthcare predictive modeling using Blockchain technology. We designed a framework, ModelChain, to integrate online machine learning with blockchains, and utilized transaction metadata for model dissemination. We also developed a new proof-of-information algorithm to determine the order of Blockchain-based online machine learning. Our next step is to evaluate ModelChain trade-offs in real-world settings such as the Patient-centered SCAlable National Network for Effectiveness Research (pSCANNER). Also, we will continue to improve the proof-of-information algorithm in terms of efficiency and scalability. We anticipate that the combination of technology and policy will be key to advance health services research and healthcare quality improvement.

**Acknowledgements**
The authors would like to thank Chun-Nan Hsu, PhD, Xiaoqian Jiang, PhD, and Shuang Wang, PhD for very helpful discussions. The authors are funded by PCORI CDRN-1306-04819. LO-M is funded by NIH U54HL108460, UL1TR001442, and VA I01HX000982. This article has been presented as a challenge-winning submission on the Office of the National Coordinator for Health Information Technology (ONC) and the National Institute of Standards and Technology (NIST) Use of Blockchain for Healthcare and Research Workshop in 2016.

# Reference

1. Navathe AS, Conway PH. Optimizing health information technology's role in enabling comparative effectiveness research. *Am J Manag Care* 2010;**16**:SP44–7.
2. Wicks P, Vaughan TE, Massagli MP, Heywood J. Accelerated clinical discovery using self-reported patient data collected online and a patient-matching algorithm. *Nat Biotechnol* 2011;**29**:411–4.
3. Grossman JM, Kushner KL, November EA, Lthpolicy PC. *Creating sustainable local health information exchanges: can barriers to stakeholder participation be overcome?*. Center for Studying Health System Change Washington, DC; 2008.
4. Wu Y, Jiang X, Kim J, Ohno-Machado L. Grid Binary LOgistic REgression (GLORE): building shared models without sharing data. *J Am Med Inform Assoc* 2012;**19**:758–64.
5. Wang S, Jiang X, Wu Y, Cui L, Cheng S, Ohno-Machado L. Expectation propagation logistic regression (explorer): distributed privacy-preserving online model learning. *J Biomed Inform* 2013;**46**:480–96.
6. Li Y, Jiang X, Wang S, Xiong H, Ohno-Machado L. VERTIcal Grid lOgistic regression (VERTIGO). *J Am Med Inform Assoc* 2015:ocv146.
7. McConaghy T, Marques R, Müller A, De Jonghe D, McConaghy T, McMullen G, et al. BigchainDB: A Scalable Blockchain Database (DRAFT) 2016.
8. Fromknecht C, Velicanu D, Yakoubov S. A Decentralized Public Key Infrastructure with Identity Retention. *IACR Cryptology ePrint Archive* 2014;**2014**:803.
9. Luu L, Narayanan V, Baweja K, Zheng C, Gilbert S, Saxena P. *SCP: a computationally-scalable Byzantine consensus protocol for blockchains*. Cryptology ePrint Archive, Report 2015/1168; 2015.
10. Greenspan G. *MultiChain Private Blockchain - White Paper*. 2015.
11. Pilkington M. Blockchain technology: principles and applications. *Research Handbook on Digital Transformations, Edited by F Xavier Olleros and Majlinda Zhegu Edward Elgar* 2016.
12. Xu X, Pautasso C, Zhu L, Gramoli V, Ponomarev A, Tran AB, et al. *The blockchain as a software connector*. 2016.
13. Lamport L, Shostak R, Pease M. The Byzantine generals problem. *ACM Trans Program Lang Syst* 1982;**4**:382–401.
14. Douceur JR. *The sybil attack*. Springer; 2002.
15. Bissias G, Ozisik AP, Levine BN, Liberatore M. *Sybil-resistant mixing for bitcoin*. ACM; 2014.
16. McConaghy T. Blockchain, Throughput, and Big Data. *Bitcoin Startups Berlin, Oct* 2014;**28.**:
17. Nakamoto S. Bitcoin: A peer-to-peer electronic cash system 2008.
18. Miller A, LaViola JJ Jr. *Anonymous byzantine consensus from moderately-hard puzzles: A model for bitcoin*. 2014.
19. Meiklejohn S, Pomarole M, Jordan G, Levchenko K, McCoy D, Voelker GM, et al. *A fistful of bitcoins: characterizing payments among men with no names*. ACM; 2013.
20. Garay J, Kiayias A, Leonardos N. *The bitcoin backbone protocol: Analysis and applications*. Springer; 2015.
21. Freund Y, Schapire RE. *A desicion-theoretic generalization of on-line learning and an application to boosting*. Springer; 1995.
22. Servedio R. Smooth Boosting and Learning with Malicious Noise. *J Mach Learn Res* 2003;**4**:633–48.
23. Lo H-Y, Chang K-W, Chen S-T, Chiang T-H, Ferng C-S, Hsieh C-J, et al. An ensemble of three classifiers for KDD cup 2009: Expanded linear model, heterogeneous boosting, and selective naıve Bayes. *JMLR W&CP* 2009;**7.**:
24. Kalai A, Kanade V. Potential-Based Agnostic Boosting. In: Bengio Y, Schuurmans D, Lafferty J, Williams CKI, Culotta A, editors. *Advances in Neural Information Processing Systems 22*. 2009. p. 880–8.
25. Freund Y, Schapire RE. *Experiments with a new boosting algorithm*. vol. 96. 1996.
26. *Health IT Standards and Health Information Interoperability | Policy Researchers & Implementers | HealthIT.gov*. n.d. URL: https://www.healthit.gov/policy-researchers-implementers/interoperability (Accessed 5 August 2016).
27. Ohno-Machado L, Agha Z, Bell DS, Dahm L, Day ME, Doctor JN, et al. pSCANNER: patient-centered Scalable National Network for Effectiveness Research. *J Am Med Inform Assoc* 2014;**21**:621–6.
28. Jones M, DuVall SL, Spuhl J, Samore MH, Nielson C, Rubin M. Identification of methicillin-resistant Staphylococcus aureus within the nation's Veterans Affairs medical centers using natural language processing. *BMC Med Inform Decis Mak* 2012;**12**:1.
29. D'Avolio L, Ferguson R, Goryachev S, Woods P, Sabin T, O'Neil J, et al. Implementation of the Department of Veterans Affairs' first point-of-care clinical trial. *J Am Med Inform Assoc* 2012;**19**:e170–6.
30. Mandel AJ, Kamerick M, Berman D, Dahm L. *University of California Research eXchange (UCReX): a federated cohort

|    | |
|----|---|
| | *discovery system*. IEEE Computer Society; 2012. |
| 31 | Jiang X, Sarwate AD, Ohno-Machado L. Privacy technology to support data sharing for comparative effectiveness research: a systematic review. *Med Care* 2013;**51**:S58. |
| 32 | Kim KK, McGraw D, Mamo L, Ohno-Machado L. Development of a privacy and security policy framework for a multistate comparative effectiveness research network. *Med Care* 2013;**51**:S66–72. |
| 33 | Gabriel SE, Normand S-LT. Getting the methods right—the foundation of patient-centered outcomes research. *N Engl J Med* 2012;**367**:787–90. |
| 34 | Frank L, Basch E, Selby JV. The PCORI perspective on patient-centered outcomes research. *JAMA* 2014;**312**:1513–4. |
| 35 | *Patient-Centered Outcomes Research*. 2013. URL: http://www.pcori.org/research-results/patient-centered-outcomes-research. |
| 36 | Washington AE, Lipstein SH. The Patient-Centered Outcomes Research Institute—promoting better information, decisions, and health. *N Engl J Med* 2011;**365**:e31. |
| 37 | Selby JV, Beal AC, Frank L. The Patient-Centered Outcomes Research Institute (PCORI) national priorities for research and initial research agenda. *JAMA* 2012;**307**:1583–4. |
| 38 | Clancy C, Collins FS. Patient-Centered Outcomes Research Institute: the intersection of science and health care. *Sci Transl Med* 2010;**2**:37cm18–37cm18. |
| 39 | Jiang W, Li P, Wang S, Wu Y, Xue M, Ohno-Machado L, *et al.* WebGLORE: a web service for Grid LOgistic REgression. *Bioinformatics* 2013;**29**:3238–40. |
| 40 | Shi H, Jiang C, Dai W, Jiang X, Tang Y, Ohno-Machado L, *et al.* Secure Multi-pArty Computation Grid LOgistic REgression (SMAC-GLORE). *BMC Med Inform Decis Mak* 2016;**16 Suppl 3**:89. |
| 41 | Wu Y, Jiang X, Wang S, Jiang W, Li P, Ohno-Machado L. Grid multi-category response logistic models. *BMC Med Inform Decis Mak* 2015;**15**:10. |
| 42 | El Emam K, Samet S, Arbuckle L, Tamblyn R, Earle C, Kantarcioglu M. A secure distributed logistic regression protocol for the detection of rare adverse drug events. *J Am Med Inform Assoc* 2013;**20**:453–61. |
| 43 | Slavkovic AB, Nardi Y, Tibbits MM. 'Secure' Logistic Regression of Horizontally and Vertically Partitioned Distributed Databases. Presented at the 2007 Seventh IEEE International Conference on Data Mining - Workshops (ICDM Workshops), Omaha, NE, USA, 28/10/2007-31/10/2007. |
| 44 | Fienberg SE, Fulp WJ, Slavkovic AB, Wrobel TA. 'Secure' Log-Linear and Logistic Regression Analysis of Distributed Databases. Presented at the International Conference on Privacy in Statistical Databases. |
| 45 | Yan F, Sundaram S, Vishwanathan S, Qi Y. Distributed autonomous online learning: Regrets and intrinsic privacy-preserving properties. *IEEE Trans Knowl Data Eng* 2013;**25**:2483–93. |
| 46 | Yu S, Fung G, Rosales R, Krishnan S, Rao RB, Dehing-Oberije C, *et al.* Privacy-Preserving Cox Regression for Survival Analysis. Presented at the Las Vegas, Nevada, USA. |
| 47 | Lu C-L, Wang S, Ji Z, Wu Y, Xiong L, Jiang X, *et al.* WebDISCO: a web service for distributed cox model learning without patient-level data sharing. *J Am Med Inform Assoc* 2015;**22**:1212–9. |
| 48 | Li H, Kadav A, Kruus E, Ungureanu C. MALT: Distributed Data-Parallelism for Existing ML Applications. Presented at the Bordeaux, France. |
| 49 | Li M, Zhou L, Yang Z, Li A, Xia F. Parameter server for distributed machine learning. *Big Learning NIPS* 2013. |
| 50 | Ho Q, Cipar J, Cui H, Kim JK, Lee S, Gibbons PB, *et al.* More Effective Distributed ML via a Stale Synchronous Parallel Parameter Server. *Adv Neural Inf Process Syst* 2013;**2013**:1223–31. |
| 51 | Li M, Andersen DG, Park JW, Smola AJ. Scaling distributed machine learning with the parameter server. *USENIX Symposium on …* 2014. |
| 52 | Dean J, Corrado G, Monga R, Chen K, Devin M, Mao M, *et al.* Large Scale Distributed Deep Networks. In: Pereira F, Burges CJC, Bottou L, Weinberger KQ, editors. *Advances in Neural Information Processing Systems 25*. Curran Associates, Inc.; 2012. p. 1223–31. |
| 53 | Chu C, Kim SK, Lin YA, Yu YY, Bradski G. Map-reduce for machine learning on multicore. *Adv Neural Inf Process Syst* 2007. |
| 54 | Boyd S, Parikh N, Chu E, Peleato B, Eckstein J. Distributed Optimization and Statistical Learning via the Alternating Direction Method of Multipliers. *Found Trends Mach Learn* 2011;**3**:1–122. |
| 55 | Dean J, Ghemawat S. MapReduce: simplified data processing on large clusters. *Commun ACM* 2008. |
| 56 | Low Y, Gonzalez JE, Kyrola A, Bickson D, Guestrin CE, Hellerstein J. GraphLab: A New Framework For Parallel Machine Learning. *arXiv [csLG]* 2014. |
| 57 | Bentov I, Lee C, Mizrahi A, Rosenfeld M. Proof of Activity: Extending Bitcoin's Proof of Work via Proof of Stake |

[Extended Abstract] y. *ACM SIGMETRICS* 2014.
58  Pass R, Tech C, Seeman L. Analysis of the Blockchain Protocol in Asynchronous Networks 2016.
59  Sompolinsky Y, Zohar A. Secure High-Rate Transaction Processing in Bitcoin. Presented at the International Conference on Financial Cryptography and Data Security.
60  Eyal I. The Miner's Dilemma.
61  Eyal I, Sirer EG. Majority Is Not Enough: Bitcoin Mining Is Vulnerable. Presented at the International Conference on Financial Cryptography and Data Security.
62  Kalodner H, Carlsten M, Ellenbogen P. An empirical study of Namecoin and lessons for decentralized namespace design. *Workshop on the* 2015.
63  Rosenfeld M. Overview of colored coins. *White Paper, Bitcoil Co Il* 2012.
64  Back A, Corallo M, Dashjr L, Friedenbach M, Maxwell G, Miller A, *et al.* Enabling blockchain innovations with pegged sidechains. *URL: Http://www Opensciencereview com/papers/123/enablingblockchain-Innovations-with-Pegged-Sidechains* 2014.
65  Bonneau J, Miller A, Clark J, Narayanan A, Kroll JA, Felten EW. SoK: Research Perspectives and Challenges for Bitcoin and Cryptocurrencies.
66  King S, Nadal S. Ppcoin: Peer-to-peer crypto-currency with proof-of-stake. *Self-Published Paper, August* 2012.
67  Bentov I, Gabizon A, Mizrahi A. Cryptocurrencies without Proof of Work. *arXiv [csCR]* 2014.
68  Stewart I. Proof of burn. bitcoin. it 2012.
69  Vukolić M. *The quest for scalable blockchain fabric: Proof-of-work vs. BFT replication*. Springer; 2015.
70  Mainelli M, Smith M. Sharing ledgers for sharing economies: an exploration of mutual distributed ledgers (aka blockchain technology). *The Journal of Financial Perspectives* 2015;**3**:38–69.
71  Walsh L, Akhmechet V, Glukhovsky M. Rethinkdb-rethinking database storage 2009.
72  Kosba A, Miller A, Shi E, Wen Z, Papamanthou C. Hawk: The blockchain model of cryptography and privacy-preserving smart contracts. *University of Maryland and Cornell University* 2015.
73  Buterin V. A next-generation smart contract and decentralized application platform. *White Paper* 2014.
74  Omohundro S. Cryptocurrencies, smart contracts, and artificial intelligence. *AI Matters* 2014;**1**:19–21.
75  Swan M. *Blockchain: Blueprint for a new economy*. ' O'Reilly Media, Inc.'; 2015.
76  Swan M. *Blockchain thinking: The brain as a dac (decentralized autonomous organization)*. 2015.
77  Szabo N. The idea of smart contracts. *Nick Szabo's Papers and Concise Tutorials* 1997.
78  Wood G. Ethereum: A secure decentralised generalised transaction ledger. *Ethereum Project Yellow Paper* 2014.
79  Swan M. Blockchain Temporality: Smart Contract Time Specifiability with Blocktime. Presented at the International Symposium on Rules and Rule Markup Languages for the Semantic Web.
80  Watanabe H, Fujimura S, Nakadaira A, Miyazaki Y, Akutsu A, Kishigami J. Blockchain Contract: Securing a Blockchain Applied to Smart Contracts. Presented at the 2016 IEEE International Conference on Consumer Electronics (ICCE), Las Vegas, NV, USA, 7/1/2016-11/1/2016.
81  Irving G, Holden J. How blockchain-timestamped protocols could improve the trustworthiness of medical science. *F1000Res* 2016;**5**:222.
82  McKernan KJ. The chloroplast genome hidden in plain sight, open access publishing and anti-fragile distributed data sources. *Mitochondrial DNA* 2015:1–2.
83  Jenkins J, Kopf J, Tran BQ, Frenchi C, Szu H. Bio-Mining for Biomarkers with a Multi-Resolution Block Chain.
84  Baxendale G. Can Blockchain Revolutionise EPRs? *ITNOW* 2016;**58**:38–9.
85  Yuan B, Lin W, McDonnell C. Blockchains and electronic health records. *Mcdonnell.mit.edu* n.d.
86  Witchey NJ. Healthcare transaction validation via blockchain proof-of-work, systems and methods. 20150332283:A1, 2015.
87  Ohno-Machado L, Bafna V, Boxwala A., Chapman BE, Chapman WW, Chaudhuri K, *et al.* iDASH. Integrating data for analysis, anonymization, and sharing. *J Am Med Inform Assoc* 2012;**19**:196–201.
88  Ohno-Machado L. To share or not to share: that is not the question. *Sci Transl Med* 2012;**4**:165cm15.
89  Fleurence RL, Curtis LH, Califf RM, Platt R, Selby JV, Brown JS. Launching PCORnet, a national patient-centered clinical research network. *J Am Med Inform Assoc* 2014;**21**:578–82.
90  Collins FS, Hudson KL, Briggs JP, Lauer MS. PCORnet: turning a dream into reality. *J Am Med Inform Assoc* 2014;**21**:576–7.